\documentclass[prd,nofootinbib,superscriptaddress,twocolumn,preprintnumbers,showpacs]{revtex4}
\usepackage{textcomp}
\usepackage{graphicx}
\usepackage{flafter}
\usepackage{natbib}
\usepackage{amsmath,amssymb}
\usepackage{bm}

\renewcommand{\vec}{\textbf}

\newcommand{\f}{\frac}
\newcommand{\p}{\partial}
\renewcommand{\d}{d}
\newcommand{\be}{\begin{equation}}
\newcommand{\ee}{\end{equation}}
\newcommand{\figref}[1]{\figurename~\ref{#1}}
\newcommand{\MS}{M_S}
\newcommand{\reff}[1]{(\ref{#1})}
\newcommand{\eref}[1]{Eq.(\ref{#1})}
\newcommand{\erefs}[1]{Eqs.(\ref{#1})}
\newcommand{\omegaob}{\bar{\omega}_0}
\newcommand{\psiob}{\bar{\psi}_0}
\newcommand{\KBM}{K}
\newcommand{\xibm}{\Lambda}

\newcommand{\etar}{\eta_r}

\begin{document}
\title{Self-collimated axial jet seeds from thin accretion disks}

\author{Giulio Tirabassi}
\email{giulio.tirabassi@upc.edu}
\affiliation{ Physics Department, ``Sapienza'' University of Rome,\\
              P.le Aldo Moro 5, 00185 Roma (Italy)}
\affiliation{ Department of Physics and Nuclear Engineering,
              Universitat Politecnica de Catalunya, Barcelona (Spain).}

\author{Giovanni Montani}
\email{giovanni.montani@frascati.enea.it}
\affiliation{ Physics Department, ``Sapienza'' University of Rome,\\
              P.le Aldo Moro 5, 00185 Roma (Italy)}       
\affiliation{ ENEA - C.R. Frascati, UTFUS-MAG (Rome, Italy).}

\author{Nakia Carlevaro}
\email{nakia.carlevaro@gmail.com}
\affiliation{ Physics Department, ``Sapienza'' University of Rome,\\
              P.le Aldo Moro 5, 00185 Roma (Italy)}


\begin{abstract}
We show how an appropriate stationary crystalline structure of the magnetic field can induce a partial fragmentation of the accretion disk, generating an axial jet seed composed of hot plasma twisted in a funnel-like structure due to the rotation of the system. The most important feature we outline is the high degree of collimation, naturally following from the basic assumptions underlying the crystalline structure. The presence of non-zero dissipative effects allows the plasma ejection throughout the axial jet seed and the predicted values of the accretion rate are in agreement with observations.
\end{abstract}
\maketitle

\section{Introduction}
A relevant and puzzling feature of compact astrophysical objects (\emph{e.g.}, Gamma Ray Burst \cite{Pi99} and Active Galactic Nuclei \cite{Kr99}) is their capability to generate highly energetic and collimated axial jets, which are well observed in the various electromagnetic bands. The way such jets remain collimated over a very long astrophysical path is a non trivial question, but indications exist that a pressure balance can take place as an effect of the medium in which they are propagating \cite{Pi99, BKL01}. However, the mechanism which is responsible for the generation of such a peculiar emission of matter and radiation remains almost unidentified. A measurement of the difficulty to find a satisfactory explanation of the jet phenomenon is provided by the fact that one of the most promising proposals postulates magnetic monopole effects \cite{BKL01, Be10, LWS87, LBC91, Lo02}. Indeed, the possibility to formulate a reliable model for the generation of matter jet (without involving exotic physics) must be regarded as a significant achievement in understanding the behavior of the accreting material near a compact astrophysical source.

An alternative framework for the investigation of the stellar wind and jet origin was offered by the studies pursued in \cite{Co05, CR06}, where it was shown how a local equilibrium configuration allows the existence of a periodic structure for the magnetic flux surfaces, as long as a proper account for the plasma magnetic backreaction is provided. This approach was extensively studied in \cite{LM10, MB11GRG, BMP11} and generalized to the global profile of the disk equilibrium in \cite{MB11}. In this scheme, the equilibrium of a rotating stellar disk could favor the emergence of wind and jet seeds already in the steady state, as discussed in \cite{MC10}. Such a work upgrades the original idea of an ideal purely rotating disk in which a crystalline magnetic field arises, by including non zero poloidal velocities and matter fluxes. This feature implies that the original local equilibrium must now deal with a non zero azimuthal and electron-momentum-balance equations. The reason that this scheme is able to account for high peaks of the vertical velocity (\emph{i.e.}, the seeds of winds and jets) consists just in the structure that the azimuthal component of the electron force balance takes in the presence of poloidal velocities. In fact, since the azimuthal electric field must identically vanish by virtue of the axial symmetry, one obtains a vertical radial velocity that contains (in its denominator) the radial component of the magnetic field (almost vanishing in the background dipole field). Since the backreaction induces an oscillating radial profile in the radial component of this field (and, in the non linear case, also in the disk mass density), it can be immediately and qualitatively recognized that, in the $O$-point of the magnetic configuration, the vertical velocity takes extremely large values. This specific picture of the disk equilibrium configuration is certainly very intriguing and promising, in view of the generation of winds or jets along the axial direction. Nonetheless, it contains three significant limitations: \emph{(i)} the analysis is performed in a local radial scenario only, which prevents a localization of the peaks within the disk; \emph{(ii)} the emerging peak array appears to have a periodic nature, in place of what is commonly observed in real sources; \emph{(iii)} the matter-flux lines are closed, since they follow the magnetic profile and therefore no real ejection of matter is possible.

The present analysis is aimed at overcoming these difficulties and, by a global study of the plasma profile in the disk, we are able to construct an essentially single peak picture (the remaining ones being strongly suppressed outward). Furthermore, by including small dissipative effects in the plasma (according to the possibility to preserve the corotation theorem \cite{BMP11}), we show how the matter-flux lines become open, allowing a real ejection of material out of the source. In this respect, our work constitutes a significant step toward a settled model for a collimated jet in the scenario introduced by Coppi \cite{Co05}. Moreover, it offers a valuable theoretical framework for facing a more phenomenological characterization of the wind or jet generation in real astrophysical sources (in stellar as well as in galactic contexts). Regarding the phenomenological impact of the model, three main merits of this study must be put in order. First of all, the collimated nature of the jet seed is guaranteed by the very short scale of the magnetic field crystalline structure in the disk (see \cite{MB11}) and by the self-consistence of the equilibrium configuration equations. Since the region in which the radial component of the magnetic field (and also the mass density) is approximately zero is very tiny, the outcoming jet is obliged to have a very small cross depth. In the scenario of periodic jet seeds discussed above \cite{MC10}, this fact is a shortcoming too, implying a small scale sequence of peaks, and therefore it is not clear how they appear when a macroscopical average is taken. Here the primary jet seed is only one and its small scale origin is directly related with its collimation. Second, it is possible to infer the location of the jet seed from the temperature profile of the disk. In fact, the latter quantity is related to a parametric polynomial whose zeros fix the position of the seed. Finally, the present model is able to account for a non-zero accretion rate, which is compatible with the typical observed values in a real system.

The paper is structured as follows. In the first Section, the basic equations and assumptions of the model for a thin accretion disk are introduced. In the second Section, the equilibrium equations are integrated and the behaviors of the mass density and vertical velocity of the disk are outlined. As a result, vertical velocity peaks, corresponding to the seed of winds and jets, are obtained. In the third Section, the consistency of the model is outlined and the temperature profile of the system is derived. In the fourth Section, we show that by introducing dissipative effects (in particular, the resistivity), the matter-flux lines become open, implying an effective ejection of material. Brief concluding remarks follow.

\section{The configuration paradigm}
In this Section, we construct an ideal stationary magneto-hydrodynamics (MHD) model for the equilibrium of a thin accretion disk, surrounding a compact astrophysical object of mass $\MS$. Such a model faces an ideal rotating plasma whose equilibrium configuration is expressed via the following MHD Navier-Stokes equation, \emph{i.e.}, the momentum conservation law: \begin{equation}
\rho \vec v \cdot \nabla \vec v = - \nabla p - \rho\nabla \chi - \vec B \times ( \nabla \times \vec B )/4\pi\;,
\label{NS}
\end{equation}
where $\rho$, $p$, and $\vec v$ are the density, pressure and velocity field of the disk, respectively, while $\chi$ is the gravitational potential of the central object and $\vec B$ is the magnetic field. The remaining equations closing the equilibrium system are the continuity equation,
\begin{equation}
\nabla\cdot(\rho\vec v)=0\;,
\label{cont}
\end{equation}
ensuring the mass conservation, and the induction equation describing the evolution of the magnetic field,
\begin{equation}
\nabla\times(\vec v\times\vec B)=0\;,
\label{frozen}
\end{equation}
corresponding to the advection of the magnetic field along the velocity flux in the plasma. Finally, we have to provide the equation of state $p=p(\rho,T)$, which characterizes the thermodynamical properties of the disk. In particular, we consider the relation $p=v_s^2\rho$, where $v_s$ denotes the sound velocity in the plasma and is assumed to be independent of $\rho$.

We now analyze the behavior of the plasma disk as emended in the background magnetic field $\vec{B}_0$, generated by the central object, and the morphology of its backreaction. In axial symmetry (with $(r,\,\phi,\,z)$ being cylindrical coordinates), the total magnetic field can be expressed as
\begin{equation}
\vec{B}=-\textbf{e}_r\,\p_r\psi/r+\textbf{e}_\phi\,I/r+\textbf{e}_z\,\p_z\psi/r\;,
\end{equation}
where $\textbf{e}_{r,\,\phi,\,z}$ denote the coordinate unit vectors and $\psi$ denotes the magnetic flux surface functions. We can split this configuration as $\psi=\psi_0+\psi_1$, where $\psi_0$ describes a dipole like contribution of the central object, \emph{i.e.},
\begin{align}\label{dipoledipole}
\psi_0=\mathcal{D}_0\,r^2\;(r^2+z^2)^{-3/2}\;, \qquad
\mathcal{D}_0=const.\;,
\end{align}
while $\psi_1=\psi_1(r,\,z)$ corresponds to the magnetic field induced by the plasma backreaction. Here, the azimuthal component is provided by the backreaction only and, therefore, the function $I$ must be regarded as a first-order term. Clearly, the backreaction magnetic function is smaller than its background counterpart, \emph{i.e.}, $|\psi_1|\ll|\psi_0|$, but its gradient (the perturbed magnetic field) can be of the same magnitude of the zeroth-order contribution. More specifically, we impose the following hierarchy in the plasma: $|\nabla\psi_1|\gtrsim|\nabla\psi_0|$ and $|\nabla^2\psi_1|\gg|\nabla^2\psi_0|$.

Our model is developed in the so-called extreme non linear regime, where $|\nabla\psi_1|\gg|\nabla\psi_0|$. In this limit, the backreaction magnetic field generated by the disk exceeds the background one due to the central object. Denoting by $k$ the average radial wave number associated to the variations of $\psi_1$ ($\p_r\psi_1\sim k\psi_1$) and because of the polynomial nature of the background field ($\p_r\psi_0\sim \psi_0/r$) the extreme non linear regime is ensured as far as we require $kr\gg1$. It has been derived in \cite{MB11} that, for a thin disk (that we are considering), the validity region for such request coincides with the whole disk. Furthermore, we prescribe that the azimuthal component of $\vec{B}_1$ satisfies the relation $|I\nabla I |\ll| \nabla\psi_1\Delta\psi_1|$. This is a natural assumption when considering backreaction profiles which are characterized by a small and sufficiently smooth azimuthal magnetic field.

In order to preserve the corotation theorem \cite{Fe37}, we consider a small enough poloidal velocity field (being of the first order), so that the advective terms in \eref{NS} can be neglected (second-order contributions) with respect to the Lorentz force. Although we will obtain high values of such poloidal velocity field, this assumption will be verified \emph{a posteriori} since it will diverge just in a narrow radial region where the magnetic force correspondingly increases. In this scheme, the equations governing the radial and vertical equilibria are
\begin{subequations}\label{bilpol1 picco}
\begin{align}
\p_r p-2\rho\omega_0\f{\d\omega_0}{\d\psi_0}r\psi_1+\f{\p_r\psi_1}{4\pi r^2}\,\Delta\psi_1=0\;,\label{bilpol1 picco a}\\
\p_{z^2} p+\rho\omega_0^2/2+\f{\p_{z^2}\psi_1}{4\pi r^2}\,\Delta\psi_1=0\;,\label{bilpol1 picco b}
\end{align}
\end{subequations}
(where $\Delta\equiv\p_r^2+\p_z^2$), while the azimuthal equilibrium equation provides only the $\phi$-component of the magnetic field and this feature is not needed for the addressed task. Thus, we reduced the model to a two-dimensional fluid dynamics problem. In \erefs{bilpol1 picco}, $\omega_0$ represents the unperturbed angular velocity assumed to be Keplerian. In fact, since the gravitational field generated by the central object is described by the Newtonian potential $\chi(r,z)=G\MS/\sqrt{r^2+z^2}$ ($G$ being the gravitational constant), the rotation is
\begin{equation}
\omega_0^{2}=\omega_K^{2}(r,z)\equiv G\MS\,(r^2+z^2)^{-3/2}\;.
\end{equation}

As already discussed, inside the disk the magnetic field of the central object is well described by the dipole like configuration given by \eref{dipoledipole}. Thus, on the equatorial plane $z=0$, the following relation hold:
\begin{equation}
\omegaob^2\;\equiv\;\omega_0^2(r,0)=
G\MS/r^3=G\MS\,\psiob^3/\mathcal{D}_0^3\;,
\label{angvelvpippo}
\end{equation}
where $\psiob\equiv\psi_0(r,0)$. Since the corotation theorem \cite{Fe37} states that the plasma angular velocity must be a function of the magnetic surfaces only, \emph{i.e.}, $\omega=\omega(\psi)$, it is natural to postulate \cite{MB11}
\begin{equation}\label{equazioneomegapsipippo}
\omega^2=G\MS\;\psi^3/\mathcal{D}_0^3\;,
\end{equation}
extending \eref{angvelvpippo} everywhere in the disk, also for a generic $\psi$ (in view of the addressed perturbative approach in which $|\psi_1|\ll|\psi_0|$). It is worth noting that, comparing \eref{equazioneomegapsipippo} and the form of $\omega_K$, the disk embedded in a dipole magnetic field of the central object can not have a Keplerian behavior far from the equatorial plane.

\begin{figure}[ht]
\centering
\includegraphics[width=\columnwidth]{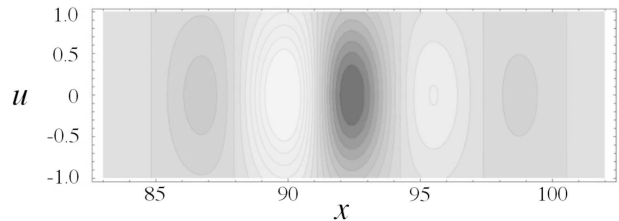}
\caption{Ideal plasma streamlines: $\theta$-function contour plot in the $(x,u)$ plane where $x=kr$ and $u=z/h$. It is clear how the crystalline structure, induced by the ring like decomposition of the backreaction magnetic field, generates a rolling pattern of plasma motion within the disk. The original streamlines generated by the central star are destroyed, except for the X-points between one roll and another, thus preventing any kind of motion between star poles through the disk.
\label{fig:flussoid}}
\end{figure}

From the mass conservation equation \reff{cont}, the plasma momentum density can be expressed via a given stream function $\theta$ (matter flux function) equivalently to the magnetic field case:
\begin{equation}\label{comparetheta}
\rho\vec v=-\textbf{e}_r\,\p_z\theta/r+\textbf{e}_\phi\,\rho\omega r+\textbf{e}_z\,\p_r\theta/r\;.
\end{equation}
The induction equation \reff{frozen} for the total magnetic field $\vec B(\psi)$ can now be restated, in the stationary limit, in terms of the functions $\theta$ and $\psi$ only, \emph{i.e.},
\begin{equation}\label{propmoto}
\text J (\theta,\psi_1)=0\;,
\end{equation}
where we have implemented the extreme non linear approximation $|\nabla\psi_1|\gg|\nabla\psi_0|$ and the Jacobian J$(\cdot,\cdot)$ is defined as
\begin{multline}
\text J(A,B)=\det\left(\begin{matrix} \p_rA & \p_{r}B \\ \p_{z}A & \p_{z}B \end{matrix} \right)=\\
=\p_rA\p_{z}B-\p_{z}A\p_rB\;.
\end{multline}
From \eref{propmoto}, we obtain $\theta=\theta(\psi_1)$, \emph{i.e.}, the motion in the meridian plane takes place along the magnetic streamlines. The plasma is therefore driven by the magnetic field and as a result, it is frozen on the plasma streamlines, preventing the plasma outflow from the system, as shown in \figref{fig:flussoid}.

\section{The emergence of a jet profile}
In order to obtain a jet like solution within the present scheme, we have to deal with a simplified form of \erefs{bilpol1 picco}. Deriving \eref{bilpol1 picco a} with respect to $z^2$ and \eref{bilpol1 picco b} to $r$, in the limit of a strong backreaction, we can obtain the following single partial differential equation (PDE):
\begin{align}
2\omega_0\f{\d\omega_0}{\d\psi_0}r \p_{z^2}(\rho\psi_1)-\f{1}{2}\,\omega_0^2\,\p_r\rho
+\f{\text J_*(\Delta\psi_1,\psi_1)}{4\pi r^2}=0\;,
\label{combo}
\end{align}
where now $\text J_*(\cdot,\cdot)$ is
\begin{multline}
\text J_*(A,B)=\det\left(\begin{matrix} \p_rA & \p_{r}B \\ \p_{z^2}A & \p_{z^2}B \end{matrix} \right)=\\
=\p_rA\p_{z^2}B-\p_{z^2}A\p_rB\;,
\label{jaco}
\end{multline}
(we note that if $A=A(B)$, then $\text J_*(A,B)=0$). This procedure is almost equivalent to taking the curl of the vectorial equation in the plane $(r,z)$. Clearly, the equation above must be retained together with one of \erefs{bilpol1 picco}. Equivalently, for each solution of \eqref{combo}, the compatibility of the original system must be checked, getting a constraint on the pressure and density profiles, \emph{i.e.}, on the sound speed too.

Let us now focus on a particular class of the $\psi_1$ functions which allows one to notably simplify \eref{combo}. In view of the paradigm defined in \cite{Co05, CR06, MB11} (where a crystalline structure of the magnetic field arises), we set
\begin{equation}\label{HE}
\Delta\psi_1=\lambda\psi_1\;,
\quad\quad \text{with $\lambda=const.<0\;$}
\end{equation}  
(commonly called the Helmoltz equation), allowing one to neglect the last term in \eref{combo}. This relation and \eref{combo} form now a consistent system.

A particular solution of \eref{HE}, approximated in the limit $z^2/h^2\ll1$, can be found as
\begin{equation}
\psi_1=\mathcal{D}_1\exp\Big[-\f{z^2}{h^2}\Big]\;\f{\sin(kr)}{g(r)}\;,
\label{psicalla}
\end{equation}
where $\mathcal{D}_1$ and $h$ are constants, and $g(r)$ is a given function such that $\p_r g(r)\sim g(r)/r$ (as we will see later, this function is connected to the temperature of the plasma). For this magnetic flux function, the eigenvalue is $\lambda=-(k^2+2/h^2)$. It is worth noting that a flux function of the form \reff{psicalla} generates, as expected, a magnetic field endowed with a crystalline structure.

Considering a generic solution of \eref{HE}, \eref{combo} reduces to a homogeneous PDE for the function $\rho$ only, since $\text J_*=0$. However, even using the form \reff{psicalla}, the resulting equation for the density of the plasma still can not be solved analytically, due to the presence of the term $\p_{z^2}(\rho\psi_1)$. Thus, in order to obtain an analytical solution, we can moreover assume
\begin{equation}
\p_{z^2}(\rho\psi_1)\sim \psi_1\,(\rho/H^2)+\rho\,(\psi_1/h^2)\;,
\label{rozzosità}
\end{equation}
where $H$ denotes the vertical scale of the disk associated to the mass profile and $h$ is the vertical scale proper of the backreaction magnetic field. In this respect, we can neglect the first or the second term on the right-hand side of the equation above, depending on whether the backreaction is contained at all or is well outside the disk, respectively.

\subsection{Behavior of the matter fluxes for $h^2\gg H^2$}
Since we are dealing with an axial jet seed of plasma which lives outside the accretion disk, and since the plasma is driven by the magnetic field only, we consider a vertical scale associated to the backreaction magnetic field much greater than the one associated to the plasma density, \emph{i.e.}, $h^2\gg H^2$. In this approximation scheme, we have $\p_{z^2}(\rho\psi_1)\simeq\psi_1\p_{z^2}\rho$.

We now take into account the thin nature of the disk $H\ll r$, where $H$ denotes, as already discussed, the vertical scale of the disk, \emph{i.e.}, the half depth. Accordingly, the relation $z\ll r$ holds. In this respect, we can approximate $\omega_0\simeq\omegaob$ and \eref{combo} is written as (considering \eref{HE})
\begin{equation}
\p_r\rho-6r^2\psi_1\,\p_{z^2}\rho/\mathcal{D}_0=0\;.
\label{combo3}
\end{equation}
\begin{figure}[ht]
\centering
\includegraphics[width=0.8\columnwidth]{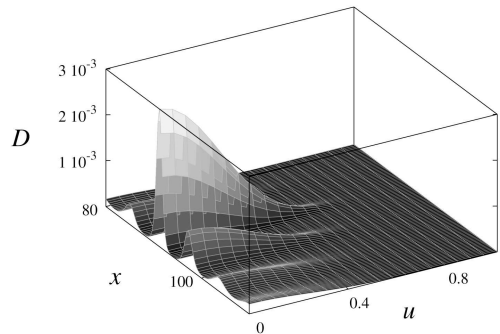}
\caption{Plot of the dimensionless density $D$ of \eref{solcombo} in the $(x,u)$ plane where $x=kr$ and $u=z/h$.
\label{fig:picco}}
\end{figure}
\begin{figure}[ht]
\includegraphics[width=0.8\columnwidth]{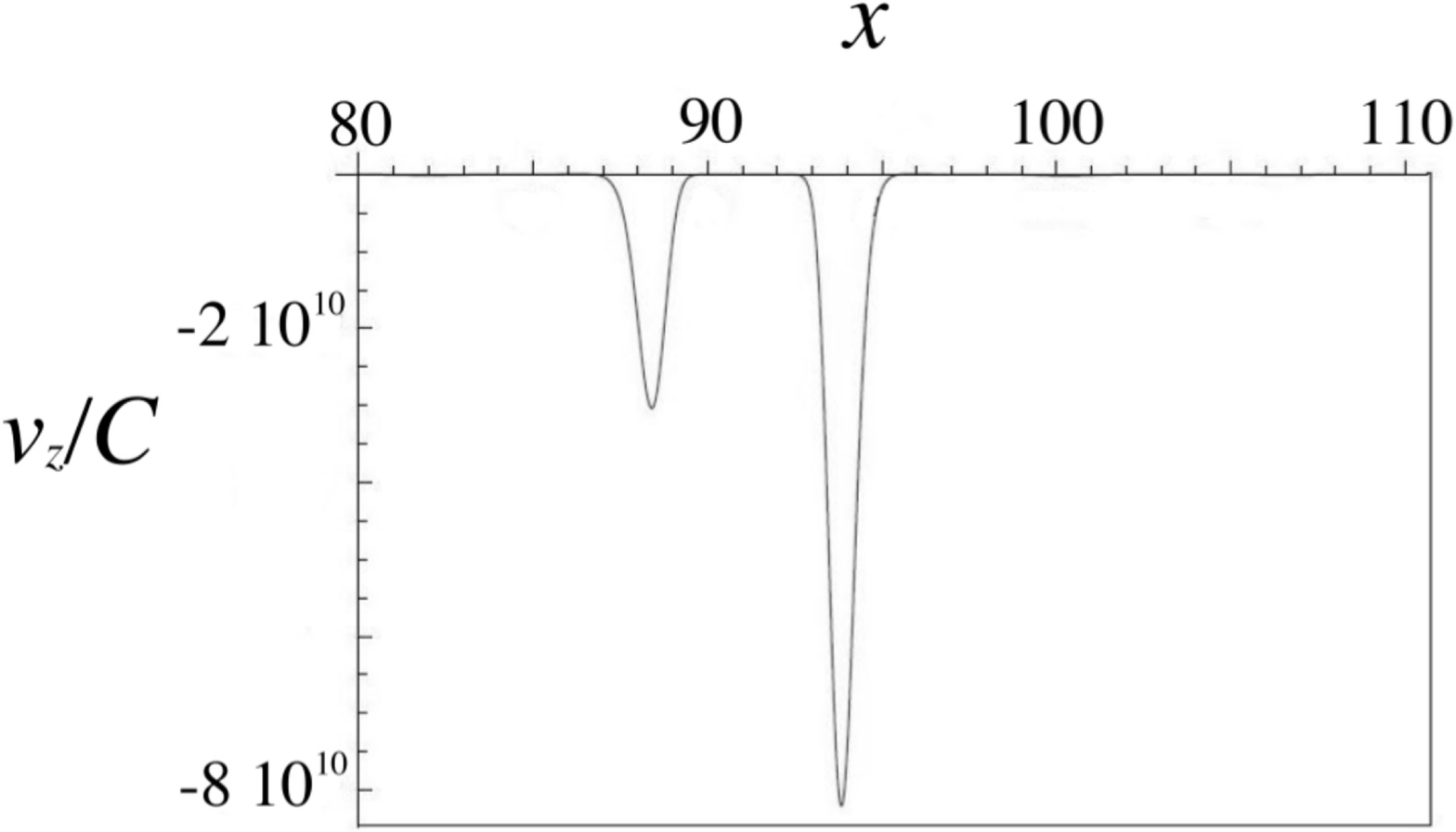}
\caption{Plot of the vertical velocity in the equatorial plane ($z=u=0$) as a function of $x=kr$ (we have set $\theta=(C/k)\psi_1$ with $C=const.$).
\label{fig:picco2}}
\end{figure}
Since the backreaction magnetic flux surface is separable in the form \reff{psicalla}, the equation above admits the following solution:
\begin{subequations}\label{rhosol}
\begin{equation}\label{solcombo}
\rho=\rho_0(r)D(r,z^2)=mr^{-d}F(z^2)N(r)\;,
\end{equation}
where $\rho_0=mr^{-d}$ (with $m=const.$ and $d=const.$; the latter parameter will be fixed from the compatibility of the system) denotes the background mass density \cite{MB11} and 
\begin{equation}\label{solcombos}
F=\exp\Big[-\frac{e^{z^2/h^2}}{6A\xi}\Big]\;,\quad
N=\exp\Big[-kr^2\f{\cos(kr)}{g(r)}\Big]\;,
\end{equation}
where (we remind the reader that $kr\gg1$) $A=\mathcal{D}_1/\mathcal{D}_0$, and
\begin{equation}
\xi=1/(k h)^2
\end{equation}
is the so-called collimation parameter. In fact, it can be easily shown that $v_r/v_z\sim\sqrt\xi$, thus, we can express the opening angle of the plasma flux as arctan($\sqrt\xi$). This way, the more $\xi$ is small, the more the flux is collimated. Furthermore, from the expression of $F$, we obtain the half thickness to be defined as $H\sim h\sqrt{6A\xi}$. 

Giving now a specific form to the function $g$, as a result it is possible to determine the explicit form of the solutions. We infer a simple third-degree polynomial,
\begin{equation}\label{g-form}
g(r)=a_3r^3+a_2r^2+a_1r\;,
\end{equation}
\end{subequations}
with $a_{1,2,3}$ constants. The choice of a polynomial is due to the previously stated request, $\p_r g\sim g/r$ and the third degree is the lower one, allowing for the presence of only one singularity in $\psi_1$, placed in the origin of the coordinate system.

\begin{figure}[ht]
\centering
\includegraphics[width=0.6\columnwidth]{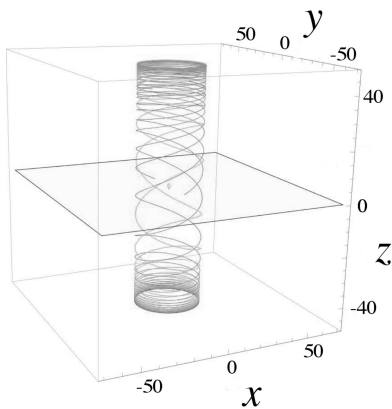}
\caption{Three-dimensional picture of the jet, in which only two different plasma trajectories are represented. The central body is pictured as a small sphere in the center of the equatorial plane $(x,y)$.
\label{fig:mbuto}}
\end{figure}

The solution of the plasma density is therefore determined and we now focus on the behavior of the vertical velocity: from \erefs{comparetheta} and \eqref{propmoto}, it is easy to recognize that
\begin{equation}
v_z\sim (\p_r\psi_1)/(\rho r)\;.
\end{equation}
In FIG.\ref{fig:picco} and FIG.\ref{fig:picco2}, we plot the dimensionless plasma density $D=\rho/\rho_0$ (on the meridian plane) and the vertical velocity (on the equatorial plane), respectively. From FIG.\ref{fig:picco}, it can be shown how the crystalline structure of the magnetic field induces a regular structure in density the profile too. In particular, denser areas followed by emptier ones are obtained, generating a fragmentation of the plasma profile. This behavior resembles the one in \cite{Co05}. In FIG.\ref{fig:picco2}, as a result, we find a space region (in correspondence to the density minima) in which the plasma develops a very large vertical velocity that can be viewed as the seed of a jet. We can extrapolate that the jet is carried outside the plasma by the magnetic field, reaching a height $h$, but this results in the motion being closed. In fact, due to the magnetic field action, the plasma is forced to move toward the equatorial plane, as shown in FIG.\ref{fig:flussoid}. Of course, this feature is in contrast with observations and we will discuss this point in the last Section. Finally, FIG.\ref{fig:mbuto} depicts a three-dimensional view of two jet streamlines, twisted together by the rotation of the accretion disk. It shows the funnel-like nature of such an axial jet and the high collimation of the stream.

\section{Thermodynamical features of the jet seed}
Let us now focus on the thermodynamic properties of the system. In particular, substituting the expressions of $\rho$ and $\psi_1$ (given by \erefs{solcombo} and \reff{psicalla}), in the system of the radial and vertical equilibrium \erefs{bilpol1 picco}, one gets a couple of PDEs for the function $p$ only. Since we are dealing with two differential equations describing a single unknown function, the compatibility of such an overdetermined system is under discussion. Introducing the following dimensionless quantities
\begin{subequations}\label{definizioni infami}
\begin{align}
x & =kr\;,\quad\qquad\;\, u=z/h\;,\quad\quad\, Y=k\psi_1/\p_r\psi_0\;,\\
\KBM & =G\MS m/\mathcal{D}_0\;,\;\;\xibm=(\KBM h)^{-2}\;,\;\; P=p/{B_{0(z)}^2}\;,\\
\tilde{P} & =2\xibm\big(P+(2\xi-1)Y^2/8\pi\big)\;,
\end{align}
\end{subequations}
and setting $d=3$ in the definition of $\rho_0$, the system \eqref{bilpol1 picco} is now rewritten as
\begin{align}\label{oddiolapressione}
\p_{x}\tilde{P}-6\xi DY=0\;,\qquad\quad
\p_{u^2}\tilde{P}+D=0\;.
\end{align}

From the second of \erefs{oddiolapressione}, we obtain the condition
\begin{equation}
\tilde{P}=-N(x)\int\!\! F(u^2)\,du^2\;,
\end{equation}
and, substituting the expression above into the first of \erefs{oddiolapressione}, the compatibility condition simply results in $\xi\ll1$, implying a well-collimated flux. The intriguing character of this issue can be immediately recognized since it has been obtained by just requiring the consistency of the model.

\begin{figure}[ht]
\centering
\includegraphics[width=0.7\columnwidth]{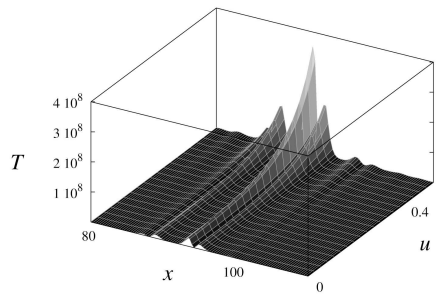}
\caption{Plot of $T=P/D$ on the $(x,u)$ plane. The parameters are set as for the previous analyses.
\label{fig:temp}}
\end{figure}

Let us now discuss the morphology of the plasma pressure obtained from $\tilde P$, and, accordingly, of the temperature $T$ of the accretion disk. In fact, these quantities are connected by the perfect gas equation of state, fixed as $p=v_s^2\rho\sim\rho T$. In FIG.\ref{fig:temp}, we plot the plasma dimensionless temperature field $T=P/D$. It is easy to realize that, in correspondence to the velocity peaks (or the density minima), the presence of temperature peaks is outlined as well. This implies that the jet seed is characterized by an upward motion of very hot plasma.   
In this respect, the temperature field can be inferred as responsible for the presence of the velocity peak itself, \emph{i.e.}, for the particular form of the $g$-function introduced in \eref{psicalla}. If the disk possesses a region in which the plasma is hotter with respect to the mean temperature, then a strong modulation of the crystalline structure appears, generating the ring decomposition and the vertical jet seed. Moreover, such an hotter region is expected, according to the observations, to belong to the inner regions of the disk. The temperature profile grows with the vertical coordinate and this is due to the fact that the jet seed is rarefied in the outer part of the system and because no cooling-by-radiation mechanism has been introduced in the model.

\section{Dissipative effects}
In the first Section, we outline how the plasma velocity (related to $\theta$) is a function of $\psi$ and as a result is parallel to the magnetic field. The motion of the jet seed is therefore closed since the particle trajectories are frozen on the closed magnetic lines. In this respect, in order to obtain a jet implying an effective matter transport outward from the disk, a mechanism responsible for the misalignment of $\vec{v}$ and $\vec{B}$ is requested. To this end, we introduce dissipative effects mainly characterized by the resistive coefficient $\etar$, since we require these effects to be relevant in the magnetic field induction equation only.
\begin{figure}[ht]
\centering
\includegraphics[width=\columnwidth]{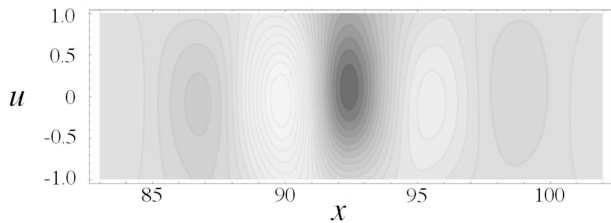}
\caption{Dissipative plasma streamlines: $\theta$-function contour plot in the $(x,u)$ plane where $x=kr$ and $u=z/h$.
\label{fig:flussovisc}}
\end{figure}
This is a regime in which resistive effects are stronger than viscous ones (indeed corresponding to the limit of a magnetic Prandtl number greater than one). Moreover, we underline that in \cite{BMP11} it has been shown how the corotation theorem \cite{Fe37} holds under the hypothesis of stationarity up to the second order in small dissipative effects, poloidal velocity and toroidal component of the magnetic field. In our approximation scheme, one obtains the following magnetic induction equation:
\begin{equation}\label{bla}
\nabla\times(\vec{v}\times\vec{B}+\etar\nabla\times\vec{B})=0\;,
\end{equation}
while the system \reff{bilpol1 picco} remains unchanged. Using the two flux functions $\theta$ and $\psi$, we can express, in the extreme non linear regime, \eref{bla} as
\begin{equation}
\text{J}(\psi_1,\theta)+\rho \etar r \Delta \psi_1=0\;,
\label{eqpolB}
\end{equation} 
in place of \eref{propmoto}. It is important to stress that the presence of a non zero constant resistivity coefficient implies the misalignment discussed above. In fact, we obtain the relation $\Delta\psi_1\times\Delta\theta=-\etar r \rho \Delta\psi_1$, which guarantees the emergence of open flux lines of matter. \eref{eqpolB} corresponds to a PDE for the $\theta$-function only, since the radial and vertical equilibrium equations remain unchanged (because of the smallness of the poloidal velocities). Thus, the functions $\psi_1$ and $\rho$ obtained in the ideal case are preserved also in the dissipative scenario. Using now
\begin{align}\label{def}
\theta(r,z)=\theta_0(\psi)+\theta_1(r,z)\;,\quad\quad
\Theta\equiv\f k{x\etar\rho}\sqrt{\xi}\;\theta_1\;,
\end{align}
(where $\theta_1$ must be an odd $z$-function in order to obtain a non-zero accretion rate of the disk), and taking the dimensionless functions \reff{definizioni infami}, \eref{eqpolB} is rewritten as
\begin{equation}\label{bruttabbestia}
\bar{\text J}(Y,\Theta)=(1-2\xi)Y\;,
\end{equation}
where $\bar{\text J}$ denotes the Jacobian determinant expressed in dimensionless coordinates. \eref{bruttabbestia} describes the $\Theta$ function only and can be solved numerically, yielding the plasma streamlines plotted in FIG.\ref{fig:flussovisc}. As discussed above, the presence of an even small constant resistivity coefficient opens the plasma streamlines, allowing an effective plasma outflow through the jet seed. The opening of the matter flux is shown in correspondence to the velocity peak of FIG.\ref{fig:picco2}.

Using now the obtained $\Theta$-function values, we can estimate the accretion rate towards the central object by
\begin{equation}\nonumber
\dot M(r)=-2\pi r \int_{-H}^{H}\!\!\!\!\!\!\!\!\rho v_r\,dz\simeq
-2\pi r \int_{-h}^{h}\!\!\!\!\!\!\!\rho v_r\,dz=4\pi\theta_1(r,h)\;,
\end{equation}
and, considering the stellar astrophysical scenario (at the boundary layer between the disk and the star), we thus obtain
\begin{equation}
\dot M\simeq 10^{-10}M_\odot/\text{Year}\;,
\end{equation}
which is a value compatible with observations \cite{BKL01}. In this estimation, we have considered a resistivity coefficient of the form $\etar\sim10^{12}(T/\text{K})^{-3/2}~\text{cm$^{-2}$ s$^{-1}$}$ \cite{Sp62} and values for mass, radius and temperature typical of the accretion disks associated to a cataclysmic variable, \emph{i.e.}, $M_{\text{disk}}=10^{-4}M_\odot$, $R_{\text{disk}}=10^6 \text{Km}$ and $T_{\text{disk}}=10^7 \text{K}$.

\begin{figure}[ht]
\centering
\includegraphics[angle=0,width=\columnwidth]{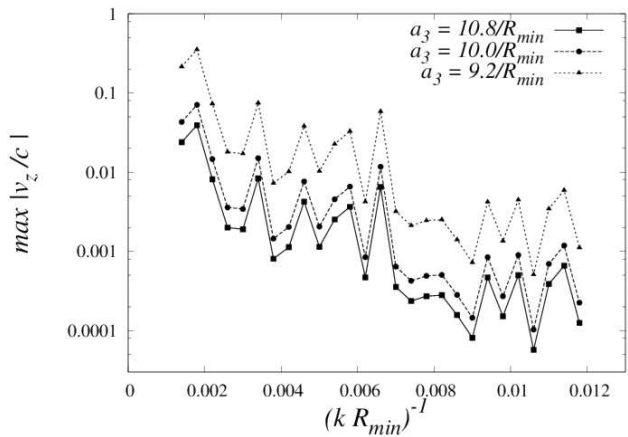}
\caption{Vertical velocities (in $c$ units) as a function of the radial length-scale of the magnetic disk perturbation for three distinct values of the parameter $a_3$ of \eref{g-form}, as indicated in the plot.
\label{fig:vzeta}}
\end{figure}

In order to get informations about the magnitude order of the outflow velocity $v_z$ through \eref{comparetheta}, we now numerically integrate \eref{bruttabbestia} with a finite-difference method. The integration is performed around the point $r=X$ corresponding to a maximum of the magnetic field perturbation in a box $[X-25/k;\,X+25/k]\times[0;\,h]$. As shown in FIG.\ref{fig:vzeta}, it is possible to obtain a broad range of vertical velocities by varying the radial length scale of the magnetic disk perturbation. In particular, for values of the parameter $a_3<10$, it is possible to obtain fast outflows compatible with observations (see, for example, \cite{Ch02} where it has been found $v_z\sim0.26c$ for the microquasar SS433 or \cite{Mu85} where $v_z\sim400-500$km/s for jets of young stars) for a good range of the perturbation length scale. We have used the following parameters: $R_{\text{min}}=10^6$m, $R_{\text{max}}=10^9$m, $\mathcal{D}_0=10^5 R_{\text{min}}^{3}$T, $\mathcal{D}_1=10^{-3}\mathcal{D}_0$, $a_1=1/R_{\text{min}}^2$, $a_2=-6/R_{\text{min}}$, $H=10^{-2}R_{\text{min}}$, $h=10H$ (\emph{i.e.}, $h\ll R_{\text{min}}$), $T_{\text{disk}}=10^7$K, $\etar\sim10^{16}(T/\text{K})^{-3/2}\text{m$^{2}$ s$^{-1}$}$, and we have fixed $X\sim3R_{min}$.

Despite the agreement with some jet observations, which is very promising for the phenomenological implementation of the model, what we can actually claim is the existence of a mechanism for the generation of a very collimated vertical matter emission, \emph{i.e.}, a jet seed or a filament (when the energy characterizing the process is rather limited). In order to deal with a real matter outflow, instead of simply a recirculation across the star magnetic field, it is necessary to include an even small amount of dissipation (viscoresistive effects), which opens the particle trajectories. Clearly, our analysis is strictly valid in the real plasma disk, say $z<H$, but its extrapolation in the region $z<h$ is still reliably possible if this part of the configuration, which we can call the magnetic disk, remains thin enough. The rather natural character of this request is outlined in FIG.\ref{fig:vzeta}, where we are considering the relation $h\sim r/30$ in the region where the jet arises. Therefore, the plots of the vertical velocity field are expected to significantly overlap those obtained from a real two-dimensional simulation of the jet, at least as far as $h\sim r$.

The only important simplification we adopted in the derivation above is that the disk is thin and its angular velocity is dependent upon the radial coordinate only. This restriction is equivalently valid in the magnetic disk region, as far as $h\ll r$, and then the idea of a highly collimated jet outflowing from the plasma disk is consistent and reliably predictive. Clearly, by this model, we can not describe the behavior of matter flux far from the equatorial plane (\emph{i.e.}, $z\gg r$) and this is the reason we can speak of jet seeds only: the word ``jet'' is well justified because of the highly collimated nature of the matter flux and the high vertical velocity, while the characterizing substantive ``seeds'' comes from the thin profile of the region where they are predictively represented here.

\section{Concluding remarks}
As in \cite{MB11} the crystalline structure of the magnetic field in a thin accretion disk (reacting to the dipole field of the central object) has been extended from a local to a global profile, in this paper we have upgraded the analysis of wind and jet seeds performed in \cite{MC10} toward a global picture. This allows one to reproduce important observational features such as the isolated nature of the jet and a non zero accretion rate of the disk as a whole. Within such a theoretical paradigm, we have developed a stationary equilibrium configuration of the disk which is able to incorporate the presence of a highly collimated jet profile, corresponding to those regions of the disk where the matter density acquires an absolute minimum in its crystalline radial dependence.

The main assumptions of our model are the relative small values of the advective forces and of the requested dissipative coefficients (with an associated Prandtl number greater than unity), according to the preservation of the corotation theorem. A non zero resistivity coefficient is required to account for real ejection of material from the disk, since the matter flux lines are no longer isomorphic to the magnetic field profile. The model contains also a free radial polynomial function $g(r)$, which fixes the position of the jet ($g=1$ would imply a jet too close to the central object, \emph{i.e.}, almost at $r=0$). However, such a function can be directly connected to the thermodynamical properties of the region where the jet arises, in particular with its temperature. The radial domain where the jet takes place corresponds to a peak of the temperature profile, which, in turns, increases with the distance from the equatorial plane.

The most appealing feature of the proposed model is that the jet collimation comes out from the compatibility of the equilibrium configuration system and, at the same time, the jet radial extension is very tiny due to the small scale of the crystalline profile. This property, associated to a non zero accretion rate of the disk (the values are in the observed range as soon as the typical orders of magnitude of a stellar system are considered), makes the jet model derived here of very promising phenomenological impact and surely an intriguing perspective in view of a non stationary extension of the equilibrium configuration. The main merit of this work is the significant upgrading provided with respect to the local analysis \cite{MC10}, which enforces the idea that the crystalline profile of the backreaction is a suitable scenario to implement the jet formation from an axisymmetric accretion structure.

{\small ** This work was partially developed within the framework of the \emph{CGW Collaboration} (www.cgwcollaboration.it). **}


\newpage
\begin{thebibliography}{0}

\newcommand{\bibi}[5]{#1, \emph{#2} \textbf{#3}, #5 (#4).}
\newcommand{\nar}{New Astr. Rev.}
\newcommand{\Apj}{ApJ}
\newcommand{\PP}{Phys. Plasma}
\newcommand{\PRE}{Phys. Rev. E}
\newcommand{\mnras}{Mon. Not. RAS}
\newcommand{\epl}{EuroPhys. Lett.}
\newcommand{\prept}{Phys. Rept.}


\bibitem{Pi99}
\bibi{T. Piran}{\prept}{314}{1999}{575}

\bibitem{Kr99}
J.H. Krolik,
\emph{Active galactic nuclei:
from the central black hole to 
the galactic environment}, 
Princeton University Press, 1999.

\bibitem{BKL01}
\bibi{G.S. Bisnovatyi-Kogan, R.V.E.  Lovelace}{\nar}{45}{2001}{663}

\bibitem{Be10}
\bibi{V.S Beskin}{Physics-Uspekhi}{53}{2010}{1199}

\bibitem{LWS87}
\bibi{R.V.V. Lovelace, J.C.L. Wang, M.E. Sulkanen M.E.}{\Apj}{315}{1987}{504}

\bibitem{LBC91}
\bibi{R.V.E. Lovelace, H.L. Berk, J. Contopoulos}{\Apj}{379}{1991}{696}

\bibitem{Lo02}
\bibi{R.V.E. Lovelace et al.}{\Apj}{572}{2002}{445}

\bibitem{Co05}
\bibi{B. Coppi}{\PP}{12}{2005}{7302}

\bibitem{CR06}
\bibi{B. Coppi, F. Rousseau}{\Apj}{641}{2006}{458}

\bibitem{LM10}
\bibi{M. Lattanzi, G. Montani}{\epl}{89}{2010}{39001}

\bibitem{MB11GRG}
\bibi{G. Montani, R. Benini}{Gen. Rel. Grav.}{43}{2011}{1121}

\bibitem{BMP11}
\bibi{R. Benini, G. Montani, J. Petitta}{\epl}{96}{2011}{19002}

\bibitem{MB11}
\bibi{G. Montani, R. Benini}{\PRE}{84}{2011}{026406}

\bibitem{MC10}
\bibi{G. Montani, N. Carlevaro}{\PRE}{82}{2010}{025402}

\bibitem{Fe37}
\bibi{V.C.A. Ferraro}{\mnras}{97}{1937}{458}

\bibitem{Sp62}
L. Spitzer Jr
\emph{Physics of Fully Ionized Gases},
Interscience, 1962.

\bibitem{Ch02}
A. Cherepashchuk,
\emph{Space Sci. Rev.} \textbf{102}(1), 23 (2002).

\bibitem{Mu85}
R. Mundt, \emph{Protostars and Planets II}, Eds. D.C. Black, M.S. Mathews, University of Arizona Press, p.414 (1985).


\end{thebibliography}
\end{document}